\documentclass[notitlepage,floats,aps,nofootinbib,preprintnumbers,superscriptaddress,twocolumn,prd]{revtex4-1}

\usepackage{amsmath}
\usepackage{amsfonts,hyperref}
\usepackage{amssymb}
\usepackage{amsthm}
\usepackage{graphicx,float}
\usepackage{verbatim}
\usepackage{color}
\usepackage[para]{threeparttable}
\usepackage{hyperref}
\hypersetup{
    colorlinks=true,
    linkcolor=blue,
    filecolor=blue,      
    urlcolor=blue,
}

\def\beq{\begin{equation*}\begin{aligned}}
\def\eeq{\end{aligned}\end{equation*}}
\def\beqn{\begin{equation}\begin{aligned}}
\def\eeqn{\end{aligned}\end{equation}}

\usepackage{amsthm}

\newtheorem{innercustomgeneric}{\customgenericname}
\providecommand{\customgenericname}{}

\begin{document}

\title{Racial Impact on Infections and Deaths due to COVID-19 in New York City}
\author{Yunseo Choi,}
\affiliation{Phillips Exeter Academy, Exeter, NH 03833, USA}
\affiliation{PRIMES USA, Massachusetts Institute of Technology, Cambridge, MA 02139, USA }
\author{James Unwin}
\email{Corresponding author: {\em unwin@uic.edu}}
\affiliation{Department of Physics,  University of Illinois at Chicago, Chicago, IL 60607, USA}

\begin{abstract}
Redlining is the discriminatory practice whereby institutions avoided investment in certain neighborhoods due to their demographics. Here we explore the lasting impacts of redlining on the spread of COVID-19 in New York City (NYC). Using data available through the Home Mortgage Disclosure Act, we construct a redlining index for each NYC census tract via a multi-level logistical model. We compare this redlining index with the  COVID-19 statistics  for each NYC Zip Code Tabulation Area. Accurate mappings of the pandemic would aid the identification of the most vulnerable areas and  permit the most effective allocation of medical resources, while reducing ethnic health disparities.
\\[4pt]
{\em Keywords: COVID-19, Redlining, Institutional Racism; Pandemic.}

\end{abstract}

\maketitle

\section{Introduction} 

Systemic racial segregation has left many United States (US) citizens---especially black Americans---cloistered in adverse living conditions. 
Broadly, institutionalized racism encompasses policies, norms, and institutional practices (both intended and unintended) that amount to racial disparity \cite{McKenzie}. Historically, institutionalized racism has left nonwhite or racially mixed communities with inadequate housing, disinvestment, and relatively low employment rates \cite{redlining_effects}. 
Many health researchers hypothesize that such practices of institutionalized racism are to blame for health disparities between ethnic groups in the US at individual and neighborhood levels \cite{redlining_health}. Moreover, such health disparities are a particular concern during the current COVID-19 pandemic.

Current efforts to quantify inequalities surrounding the COVID-19 pandemic in the US (see e.g.~\cite{Garg,Raifman}) rely on identifying the vulnerability of subgroups according to traditional CDC-defined risk factors such as old age and underlying conditions \cite{CDC}. However, racial differences in the number of COVID-19 cases and deaths are so severe that traditional risk factors alone cannot fully explain such disparity \cite{Yancy,Laurencin,Khmaissia,Coven,Schmitt,Almagro,Hooper}. In this study, we show that in New York City (NYC) the demographics of a neighborhood can imply enhanced risk for its residents and  should be considered when measuring an individual's vulnerability to COVID-19, in addition to the CDC traditionally defined risk factors. While several studies make use of the preexisting health surveys to arrive at their results, we make use of the data from the actual spread of the disease in New York City to arrive at our conclusions.  For other COVID-19 studies focused on NYC, see e.g.~\cite{Petrilli,Wadhera,Borjas,Coven,Schmitt,Almagro,Khmaissia,hospitalization}.

\begin{figure}[H]
 \centering
\includegraphics[width=0.9\linewidth]{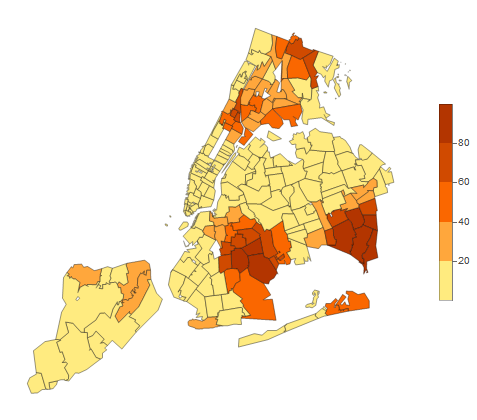}
    \vspace{-4mm}
   \caption{\hbox{Percentage of black residents in each NYC zip code~\cite{census}.} \label{F1}}
\end{figure}

Specifically, here we compare COVID-19 data to a ``redlining'' index we construct for New York City (NYC). The term "redlining" refers to discriminatory practices in which banks historically avoided investments based on neighborhood demographics: therefore, denying services to specific ethnic groups based on the locations of their residences \cite{hmda_def}. 
Historically, banks disproportionately denied mortgage applications from black Americans, barring them from entering more affluent, traditionally white communities. 
Such practices have been a real and significant detriment to black Americans. In the context of health research, redlining and other mortgage discrimination have been empirically blamed for racial health disparities as such practices would assign black Americans to poor neighborhoods with lower standards of living.  Limited access to nearby health care, poor air and water quality, and stress from high levels of crime and impoverishment mean that living standards can be closely linked with to health levels in the community \cite{redlining_health}. Figure~\ref{F1} shows the distribution of black residents in NYC.
 
This paper is structured as follows:  In Section \ref{S2}, we outline the construction of a redlining index for each census tract. Then in Section \ref{S3}, we discuss the COVID-19 statistics for NYC and compare these to the redlining index of Section \ref{S2}. In Section \ref{S5}, we discuss certain limitations of our model  and possible extensions, and in Section \ref{S6}, we highlight the significance of our findings. 

\newpage
\vspace{-2mm}
\section{Redlining Index}
\label{S2}
\vspace{-2mm}

To assuage the public's concern on mortgage discrimination, since 1975, the Federal Reserve Board has made it compulsory for financial institutions to release information about the mortgage applicants and their applications through the Home Mortgage Disclosure Act (HMDA) \cite{hmda_def}. Currently, this data is publicly available online \cite{HMDA}.
However, only a few researchers to-date have made use of the HMDA database in the context of health research. In these existing studies, the impacts of redlining on long-term, noncommunicable diseases such as cancer and those relating to perinatal health have been studied  \cite{Beyer, Mendez,Bemanian}. Another study \cite{Gee} explored the effects of redlining on access to medical resources. These studies concluded that redlining has a statistically significant influence in increasing the rates of noncommunicable diseases and in decreasing access to healthcare.  

Here we examine the relationship between redlining and COVID-19 infections and outcomes. As such, we also present the first study of the impact of redlining on the spread of communicable diseases.  To construct a redlining index we follow similar method to that in \cite{Beyer, Mendez,Bemanian} and make use of the publicly available HMDA  data sets for years 2013-2017  \cite{HMDA}.  In these data sets, information about the applicant such as the applicant's ethnicity, income, loan amount, and sex was reported. Information about the application, which includes the purpose of the mortgage and the property type, was also reported. The smallest unit of neighborhood reported in the HMDA data set is the census tract.

Since we are interested in the health disparities  between black and white ethnic groups,  we excluded primary applicants that did not identify as black or white. We also excluded applications for multi-family housing or home improvement purposes, as well as incomplete and withdrawn applications, from our analysis. After this filtering, there was a total of $208,960$ applications accounted for across $2095$ census tracts\footnote{There are $2168$ census tracts within NYC, those with very small or no population or no population where omitted. With this omission the analysis still covers all $177$ Zip Codes of NYC.} within the five year span of 2013-2017. We then geocoded the census tracts into Zip Code Tabulation Areas (ZCTA) using the Census Bureau's Relationship File \cite{USCB2}.

Using the HMDA data, we constructed a redlining index using a multilevel logistical model and then evaluated it on each census tract in NYC. The main predictor of the logistical model was the ethnicity of the primary applicant. The outcome to be measured was the log-odds of the probability of mortgage acceptance $p_{ij}$, where  $j$ indexes each census tract and $i$ indicates each individual within census tract $j$. Two covariates were utilysed, based on the variables shown to be influential in previous studies \cite{Beyer, Mendez,Bemanian}:  the applicant's sex and the ratio between the amount of loan requested given their income. 

The index was computed from the two-level equations: 
\beq
{\rm Level~1}:&\\
    \log &[p_{ij}/(1-p_{ij})] = \beta_{0j} + \beta_{1j} r_{ij} + \beta_{2j}s_{ij} + \beta_{3j}l_{ij}\\[4pt]
{\rm Level~2}:&\\
    \beta_{kj} &= \gamma_{k0} + u_{kj} \quad\text{ for}~k \in \{0, 1,2,3\}
\eeq
where, $r$ and $s$ are the ethnicity and sex of an applicant $i$ in census tract $j$ (with $r_{ij}=1$ for white and $r_{ij}=0$ for black; $s_{ij}=1$ for a male and $s_{ij}=1$ for a female), and  where $l$ is the loan to income ratio of the applicant. 

In level 2, the coefficients $\beta_{kj}$ are then identified with a fixed factor  $\gamma_{k0}$, the coefficient that best fits all of the data points, and a variation between census tracts $j$ captured by $u_{kj}$, with an assigned value such that each $u_{kj}$ best fits all of the data points within census tract $j$. 

\begin{figure}[t]
\vspace{-5mm}
 \centering
 \includegraphics[width=0.9\linewidth]{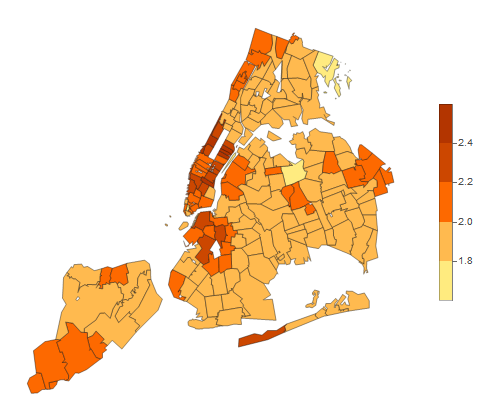}
\vspace{-5mm}
 \caption{Values of the Redlining Index $R$ constructed here. \label{F5}}
\end{figure}

Notably, $\beta_{1j}$, which tracks the ethnicity of the applicants, provides a measure of the black-to-white difference in mortgage acceptance for census tract $j$. Each of the four variations $\beta_{kj}$ (with $k\in0,1,2,3$) were tested on whether they improve the fit in terms of the $\chi^{2}$ statistic. Permitting for variations due to the sex of the applicant, $u_{2j}$, was shown to not improve the fit, and therefore, was excluded from the final model. However, the fixed effect on the sex of the applicant $\gamma_{20}$ was retained. 

From fits of the logistical model to the HMDA data we constructed the redlining index: $R=e^{\beta_{1j}}$ and quantified each census tract along a continuous scale of mortgage loan discrimination. In addition to this, one can identify the global component of the redlining index $R_F=e^{\gamma_{k0}}$ such that $R=R_F e^{u_{kj}}$.
For the 5-year dataset analysed, the index $R$ took values in the range $1.70$ to $2.48$ over the 177 NYC ZCTA. For reference, $R=2.0$ implies the probability of mortgage acceptance of a white individual is twice that of a black individual in a given census tract (adjusting for sex and loan to income ratio).  An average of $99.7$ applications were considered from each census tract, and an average of $1180.6$ applications were considered for each ZCTA.  The percentage of mortgage denial from 2013-2017 in NYC ranged from $19.6 \%$ to $26.7 \%$.

    Table \ref{fig:odds_ratio} displays the the adjusted likelihood ratio of loan acceptance between 2013-2017 after each applicant was adjusted for their loan to income ratio and their sex. In addition to calculating the index $R$ from all of the data from 2013-2017, an unadjusted (for sex or income) index $R^U$ with $\beta_{2j}=\beta_{2j}=0$ was also calculated for individual years. For reference,  in the unadjusted case the global component of the index was found to be $R^U_F= 2.15$ for 2013  and $R^U_F=2.09$ for 2017. In both the adjusted index $R$ and unadjusted index $R^U$, a white applicant was about twice as likely to have their loan accepted than a black applicant in each of the years spanning 2013-2017.

\begin{table}[t]
\centering
\vspace{-2mm}
\begin{tabular}{|c|c|c|c|c|c|}
 \hline
  & \multicolumn{4}{c|}{Applicant Race} & \\
  \cline{2-5}
  & \multicolumn{2}{c|}{Black} & \multicolumn{2}{c|}{White} & \\
  \cline{2-5}
  Year & \# & \% denied & \# & \% denied & Global redlining index ($R_F$) \\
  \hline
  2013 & 9930 & 40.2 & 46475 & 23.8 & 1.88 (1.77, 1.99) \\
  \hline
  2014 & 7203 & 37.8 & 29848 & 23.4 & 1.93 (1.81, 2.01) \\
  \hline
  2015 & 7487 & 34.8 & 32249 & 20.8 & 1.95 (1.83, 2.07)\\
  \hline
  2016 & 8090 & 37.1 & 32930 & 20.6 & 2.19 (2,06, 2.33)\\
  \hline
  2017 & 7200 & 29.9 & 27548 & 17.0 & 2.06 (1.92, 2.22) \\
  \hline
\end{tabular}
\vspace{-2mm}
    \caption{    \label{fig:odds_ratio}   Adjusted Odds Ratio 2013-2017 of Loan Acceptance, brackets show the 95\% confidence interval.}
\vspace{-5mm}
\end{table}

After the redlining indices were calculated for each census tract, we geocoded the census tracts into ZCTA. We then weighed each census tract by their population and calculated the redlining index for each ZCTA. The results are illustrated in  Figure \ref{F5}. Higher index scores indicate predominantly white, more affluent areas.  Neighborhoods with the highest indices were Upper West $(2.33)$ and Upper East $(2.31)$, and those with the lowest scores were Rockaways $(1.86)$ and Southeast Bronx $(1.88)$. 

    \vspace{-3mm}
\section{COVID-19 Spread in NYC}
    \vspace{-3mm}
\label{S3}

COVID-19 is an infectious disease caused  by the severe acute respiratory syndrome coronavirus 2 (SARS-CoV-2). The pathogen was first identified in Wuhan, China in December 2019 and rapidly  led to a worldwide pandemic, this was particularly pronounced in the US. Our analysis focuses on the spread of COVID-19 in NYC, which was an early epicentre for the pandemic in 2020.

\begin{figure}[H]
 \centering
\includegraphics[width=0.9\linewidth]{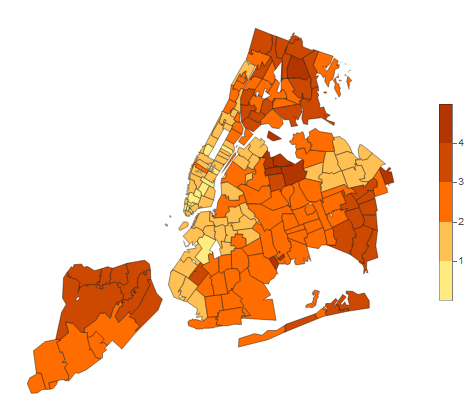}
    \vspace{-4mm}
   \caption{Number of COVID-19  cases per 100 residents. \label{F2}}
\end{figure}

\begin{figure}[H]
 \centering
 \includegraphics[width=0.9\linewidth]{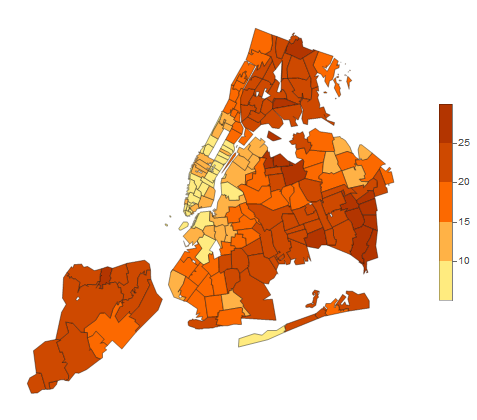}
\vspace{-4mm}
 \caption{Percentage of positive COVID-19  tests.  \label{F3}}
\end{figure}

\begin{figure}[H]
 \centering
 \vspace{-1mm}
\includegraphics[width=0.9\linewidth]{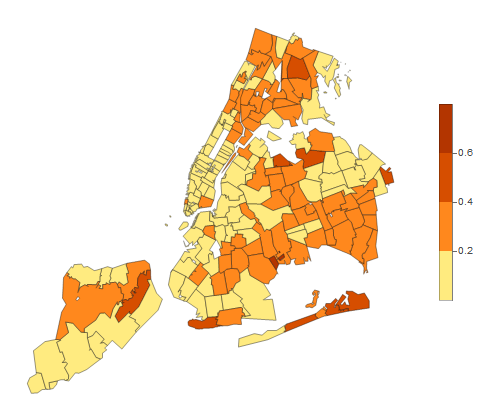}
\vspace{-4mm}
 \caption{Number of COVID-19 deaths per 100 residents. \label{F4}}
\end{figure}

Data on the spread of COVID-19 was retrieved from NYC's official website \cite{NY}, which was updated daily.\footnote{Specifically, two data sets were used: 
`tests-by-zcta', detailing the number of total tests and the number of positive tests across the 177 NYC ZCTA since April 1st; and
`data-by-modzcta', detailing the number of deaths in each ZCTA since May 18th.} 
Tests and cases with unknown ZCTA were excluded from our analysis. As of $5/26/2020$, the ZCTA of $3.1\%$ of all positive tests and that of $1.1\%$ of all tests could not be identified.   We assigned ZCTA to neighborhoods using the `Zip Code Definitions of NYC Neighborhoods'~\cite{Zip}. In Figures~\ref{F2}-\ref{F4} we illustrate the variation in the number of confirmed cases, percentage of positive tests, and total deaths resulting from COVID-19 as of June 30 2020.

    \vspace{-4mm}

\vspace{-3mm}

\begin{figure*}[t]
 \includegraphics[width=0.6\columnwidth]{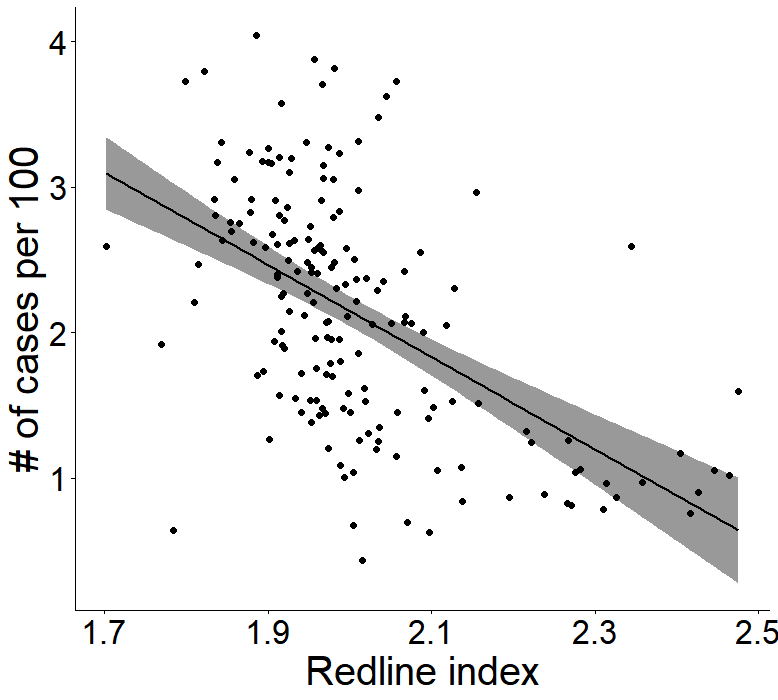}~~
 \includegraphics[width=0.6\columnwidth]{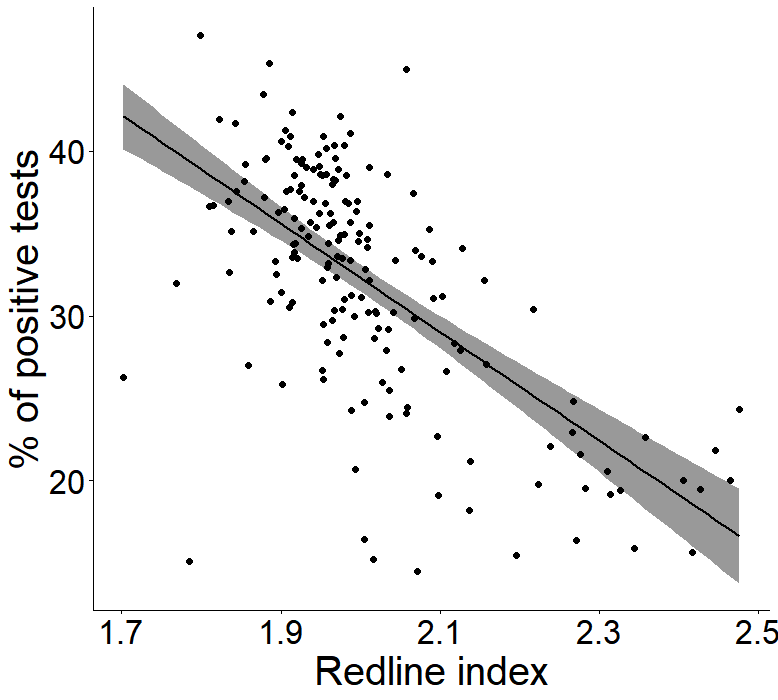}~~
 \includegraphics[width=0.6\columnwidth]{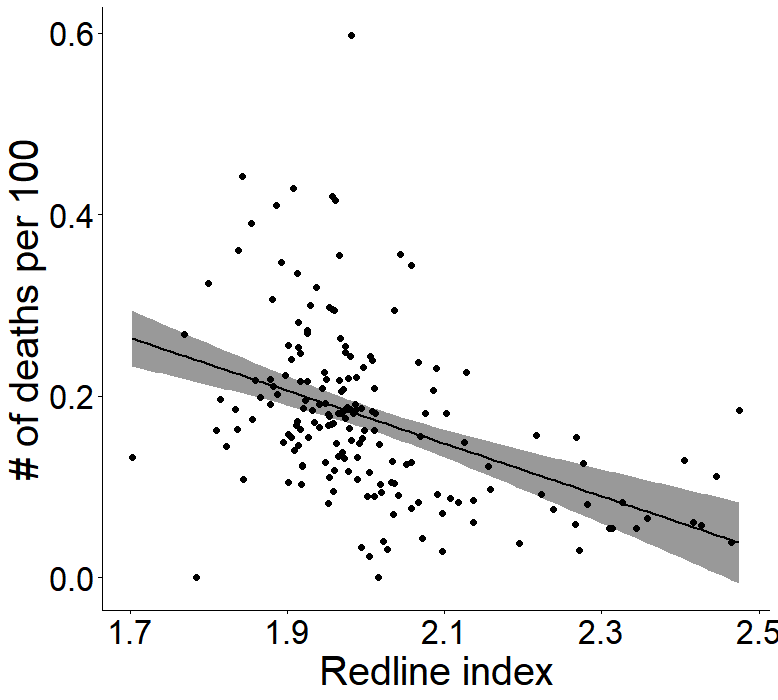}
 \caption{Scatterplots exploring correlations between the redlining index $R$ and COVID data between 15th - 20th of May 2020:  
 {\em Left.}~$R$ vs  the number of confirmed COVID-19 cases; {\em Centre.}~$R$ vs the percentage of positive tests. {\em Right.}~$R$ vs the number of COVID-19 deaths. A Best fit line is shown, with the $95\%$ confidence interval (shaded region).
  \label{F7}}
\end{figure*}

\begin{figure}[b]
\vspace{-5mm}
 \includegraphics[width=0.99\columnwidth]{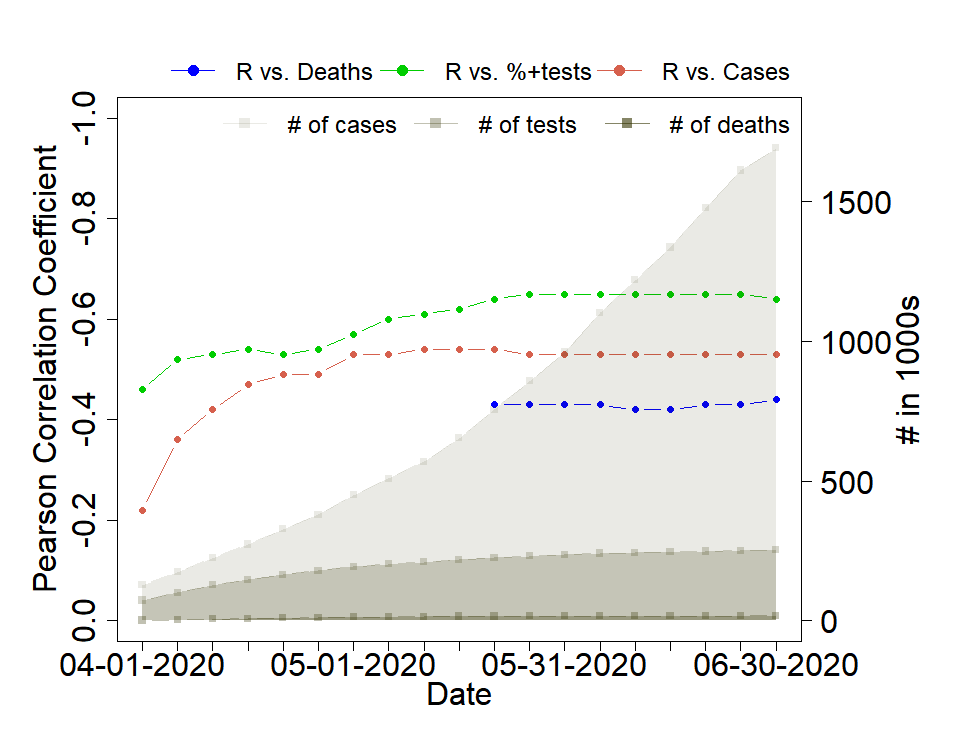}
\vspace{-5mm} \caption{Time evolution of the correlation coefficients comparing $R$ to the three NYC COVID data sets: confirmed cases, percentage positive tests, and number of deaths. The number of cases, tests, deaths are shown shaded (with RH axis). \label{F7a}}
\end{figure}

The predominantly white neighborhoods of Greenwich Village/Soho reported $0.89$ cases per 100 residents ({\em phr}), and lower manhattan had 0.078 {\em phr}, the lowest infection numbers in NYC. While West Queens and Rockaways were among the highest numbers of cases ($3.52$ and $3.51$ {\em phr}, respectively), and both also had very low redlining scores ($R=1.91,~1.86$).  Similar statements hold for the proportion of positive tests. Moreover, Greenwich Village/Soho had the least COVID deaths ($0.065$ {\em phr}), whilst Rockaways which is predominantly Black/Latino, reported the most deaths ($0.46$~{\em phr}). 

Using the redlining index constructed in Section \ref{S2}, we compute the Pearson correlation coefficient \cite{Pearson} between the redlining index $R$ and three COVID-19 data sets:
\begin{itemize}
\item The number confirmed COVID infections case.
\item The percentage of positive tests `$\%+$test'.
\item The number of COVID deaths.
\end{itemize}
 Specifically, we computed the correlation coefficient over five day periods starting from April 1$^{\text{st}}$ 2020 until June 30$^{\text{st}}$ 2020.  Over this 90 day period, the cumulative number of cases increased from 73,533 to 252,585, the cumulative  number of tests rose  from 127,550 to 1,691,978, and the number of deaths spiked from 1374 to  18,492.

To gain some intuition, we took the data for each 5-day period for which we calculated the correlation coefficient and mapped each ZCTA to points on scatterplots in the planes of $R$ versus cases, percentage of positive tests, and deaths. We show one set of plots for 15th - 20th of May 2020 in Figure~\ref{F7}. Moreover, Figure \ref{F7a}  shows the evolution of the  correlation coefficients over time (and a table of the coefficients is given in the appendix, along with the associated $p$ values). Inspecting Figure~\ref{F7a}, we note that while the correlations started out relatively weak, they all significantly strengthened over time and also settled.

It can be seen that the redlining index establishes a significant negative correlation with all three COVID data sets. This suggests that in redlined neighborhoods, the containment of the disease is harder, likely  due to a lack of sufficient medical resources. It may also suggest that less individuals tend to seek medical care, and perhaps that more individuals have to work and thus are at risk of infection. Regardless the reason, the results emphasize the need for more medical resources in redlined areas. 

\section{Discussion}
\label{S5}

Since COVID-19 data was only reported in NYC in each ZCTA, in our analysis we worked at the ZCTA-level. However, had COVID-19 data been released at the census tract level, a more detailed analysis could have been conducted. Such a fine detailed analysis would be interesting for identifying sub-pockets of vulnerable individuals. 
Furthermore, while the HMDA data does significantly increase the transparency of mortgage discrimination, potentially critical information such as the applicant's employment status, debt, and credit score were not reported. As such, these factors could not be included. 

Notably, a previous study on the 1993-1999 HMDA data set \cite{Dietrich} observed that an application from a black applicant was more likely to have missing ethnicity information than that of a white applicant. The study thus concluded that mortgage discrimination visible in the HMDA database is an underestimation of the true severity of the problem. This may imply that redlining may have an even larger impact on racial health disparities than found in our analysis leading to even stronger correlations than reported in Figure \ref{F7a}.

Moreover, although we focused on black and white ethnic groups in this study,  a potential future research directions would be to observe the impacts of residential segregation on the spread of pandemic on other ethnic groups. In particular, previous studies have concluded that the healthiness of the Hispanics in Milwaukee \cite{Bemanian} and that of Chinese Americans \cite{Gee} in Los Angeles are affected by residential segregation. 

\vspace{-2mm}
\section{Conclusion}
\label{S6}
\vspace{-1mm}

Our analysis quantifies the impact of the COVID-19 pandemic on black Americans, a sub-group which has previously shown to be  disproportionately affected by the pandemic  \cite{Coven,Schmitt,Almagro,Hooper,Khmaissia,Laurencin, Yancy}.  As of June 2020, the rate of confirmed COVID-19 cases among black NYC residents were roughly $60\%$ higher that of the white population after age adjustment \cite{NY} and the number of COVID related deaths was double for black Americans, compared to white Americans. The risk factors determined by the CDC \cite{CDC}---old age and various underlying conditions---alone are not enough to explain such disparity. This naturally raises the question about whether medical resources were distributed equally among neighborhoods or whether certain subgroups are more or less likely to reach out of medical assistance. This study has endeavored to address these apparent  health disparities through the lens of historical residential segregation. 

Moreover, this work contributes to measuring the lasting impacts of institutionalized racism on the spread of communicable diseases (taking COVID-19 as a prime example). While the medical literature is clear that environmental factors influence healthiness, very few studies have quantified residential segregation and measured its relationship with racial health disparities and even those that have primarily examined exclusively noncommunicable diseases   \cite{Beyer, Mendez,Bemanian,Gee}. Notably, reasons for why neighborhood-level factors should influence healthiness---such as stress and scarce medical resources---apply similarly to communicable and noncommunicable diseases.

This work aims to minimize racial health disparities as a consequence of the lasting impacts of institutionalized racism, specifically during a pandemic when such disparities are amplified. As demonstrated by the present case study of NYC, we suggest that such index based analyses may be helpful in predicting the vulnerability of subgroups in other cities that the COVID-19 has yet to hit and to prepare for future pandemics. 

Accurate mappings of this pandemic allow us to predict the spread of a communicable diseases and identify the most vulnerable subgroups. This information should be acted upon to more appropriately allocate medical resources in the future, to support the communities and neighborhoods that are the most in need. Ultimately, an accurate model of the spread of the COVID-19 can minimize the lasting impacts of institutionalized racism and ensure that ethnicity is not what guarantees good healthcare. In the long run, quantitative analyses, such as presented here, can guide policies to aid in the reduction of health disparities in the post-COVID-19 era.

\vspace{3mm}
{\bf Acknowledgements.}
This research was undertaken as part of the MIT-PRIMES program.

\section*{Appendix: Tables of correlation coefficients}

This appendix provides a tabulation in Table \ref{T2} of the Pearson correlation coefficients calculated in Section \ref{S3}. These tabulated results are presented graphically in Figure \ref{F7a}  of the main text. Note that deaths were only recorded in each ZCTA from May 18th. The table also indicates the associated $p$-values for each time period. Observe that while for the earliest date (4/1/2020) the $p$-value was of marginal significance ($p< 0.01$), for subsequent dates the correlation was found to be pronounced between all quantities with $p< 0.001$.

\begin{table}[b!]
 \begin{tablenotes}
   \centering
    \begin{tabular}{|c|c|c|c|}
    \hline
         Date & Cases & $\%+$tests & Deaths  \\
    \hline
    4/1 & -0.22\tnote{*} &-0.46\tnote{**}& -\\
    \hline
    4/6 & -0.36\tnote{**} &-0.52\tnote{**}& -\\
    \hline
    4/11 & -0.42\tnote{**}& -0.53\tnote{**}& -\\
    \hline
    4/16 & -0.47\tnote{**}& -0.54\tnote{**}& -\\
    \hline
    4/21 & -0.49\tnote{**}& -0.53\tnote{**}&-\\
    \hline
    4/26 & -0.49\tnote{**}& -0.54\tnote{**}& -\\
    \hline
    5/1 &-0.53\tnote{**}& -0.57\tnote{**}&-\\
    \hline
    5/6 & -0.53\tnote{**}& -0.60\tnote{**}&-\\
    \hline
    5/11 & -0.54\tnote{**}& -0.61\tnote{**}&-\\
    \hline
    5/16& -0.54\tnote{**}& -0.64\tnote{**}&-\\
    \hline
    5/21& -0.54\tnote{**}& -0.64\tnote{**}&-0.43\tnote{**}\\
    \hline
    5/26& -0.53\tnote{**}& -0.65\tnote{**}&-0.43\tnote{**}\\
    \hline
    5/31& -0.53\tnote{**}& -0.65\tnote{**}&-0.43\tnote{**}\\
    \hline
    6/5& -0.53\tnote{**}& -0.65\tnote{**}&-0.43\tnote{**}\\
    \hline
    6/10& -0.53\tnote{**}& -0.65\tnote{**}&-0.42\tnote{**}\\
    \hline
    6/15& -0.53\tnote{**}& -0.65\tnote{**}&-0.42\tnote{**}\\
    \hline
    6/20& -0.53\tnote{**}& -0.65&\tnote{**}-0.43\tnote{**}\\
    \hline
    6/25& -0.53\tnote{**}& -0.65\tnote{**}&-0.43\tnote{**}\\
    \hline
    6/30& -0.53\tnote{**}& -0.64\tnote{**}&-0.44\tnote{**}\\
    \hline
    \end{tabular}
    \caption{$*$ $p$-value $< 0.01$ ; $**$ $p$-value $< 0.001$.}
    \label{T2}
\end{tablenotes}
\end{table}

\end{document}